\documentclass{aa}  

\usepackage{graphicx}

\usepackage{txfonts}
\usepackage[dvipsnames]{color}
\usepackage{soul}
\newcommand{\LCDM}{$\Lambda$CDM}
\newcommand{\kms}{$\mathrm{km} \, \mathrm{s}^{-1}$}
\newcommand{\Vhalo}{$V_\mathrm{h,max}$}
\newcommand{\vhalo}{$V_\mathrm{h}$}
\newcommand{\Vrot}{$V_\mathrm{rot,HI}$}
\newcommand{\vrot}{$V_\mathrm{rot}$}
\newcommand{\micron}{$\mu$m}
\newcommand{\Rout}{$R_\mathrm{out,HI}$}
\newcommand{\rout}{$R_\mathrm{out}$}
\newcommand{\Vout}{$V_\mathrm{out,HI}$}

\newcommand{\mstar}{$M_\ast$}
\newcommand{\Mhalo}{$M_\mathrm{h,200c}$}
\newcommand{\mhalo}{$M_\mathrm{h}$}

\begin{document}

   \title{An assessment of the ``too big to fail'' problem for field dwarf galaxies in view of baryonic feedback effects}


   \author{E. Papastergis
          \inst{1}\fnmsep\thanks{\textit{NOVA} postdoctoral fellow}
          \and
          F. Shankar\inst{2}
          }

   \institute{Kapteyn Astronomical Institute, University of Groningen,
              Landleven 12, Groningen NL-9747AD, The Netherlands\\
              \email{papastergis@astro.rug.nl}
         \and
             School of Physics and Astronomy, University of Southampton, Southampton SO17 1BJ, UK\\
             \email{f.shankar@soton.ac.uk}
             }

    \titlerunning{TBTF problem versus baryonic feedback}
    \authorrunning{E. Papastergis \& F. Shankar}


 
  \abstract
   { 
   Recent studies have established that extreme dwarf galaxies --whether satellites or field objects-- suffer from the so called ``too big to fail'' (TBTF) problem. Put simply, the TBTF problem consists of the fact that it is difficult to explain both the measured kinematics of dwarfs and their observed number density within the \LCDM \ framework. The most popular proposed solutions to the problem involve baryonic feedback processes. For example, reionization and baryon depletion can decrease the abundance of halos that are expected to host dwarf galaxies. Moreover, feedback related to star formation can alter the dark matter density profile in the central regions of low-mass halos. In this article we assess the TBTF problem for field dwarfs, taking explicitly into account the baryonic effects mentioned above. We find that 1) reionization feedback cannot resolve the TBTF problem on its own, because the halos in question are too massive to be affected by it, and that 2) the degree to which profile modification can be invoked as a solution to the TBTF problem depends on the radius at which galactic kinematics are measured. Based on a literature sample of $\sim$90 dwarfs with interferometric observations in the 21cm line of atomic hydrogen (HI), we conclude that the TBTF problem persists despite baryonic effects. However, the preceding statement assumes that the sample under consideration is representative of the general population of field dwarfs. In addition, the unexplained excess of dwarf galaxies in \LCDM \ could be as small as a factor of $\approx 1.8$, given the current uncertainties in the measurement of the galactic velocity function. Both of these caveats highlight the importance of upcoming uniform surveys with HI interferometers for advancing our understanding of the issue. 
   }


   \maketitle
%

\section{Introduction}

The lambda cold dark matter (\LCDM) cosmological model has been long established as the prevailing paradigm for cosmology and extragalactic astronomy. Even though \LCDM \ is extremely successful at reproducing the large-scale characteristics of our universe \citep[e.g.,][]{PlanckXVI2014,Samushia2013}, there remain a number of observational challenges to the model at the small scales associated with the formation of dwarf galaxies. One of the most pressing issues is the ``too big to fail'' (TBTF) problem, first identified in the population of bright Milky Way (MW) satellites by \citet{Boylan2011,Boylan2012}. In simple terms, the TBTF problem refers to the fact that the observed kinematics of bright MW satellites imply that they are hosted by low-mass dark matter (DM) subhalos. However, a MW-sized halo is expected to have many more such subhalos in \LCDM \ than the number of bright satellites actually orbiting our galaxy.

Several possible solutions to the TBTF problem in the MW context have been put forward, related for example to the uncertainty in the mass of the halo hosting our Galaxy or pointing out the statistical weakness  of this ``sample of one'' challenge \citep[e.g.,][]{Wang2012,Vera2013,PurcellZentner2012}. However, subsequent observational results have cast these specific solutions into doubt. In particular, the TBTF problem seems to be present not only in the satellite system of the MW, but also in that of Andromeda \citep{Tollerud2014}. In addition, the TBTF problem does not only concern satellites but also field dwarf galaxies in the Local Group and Local Volume \citep{Ferrero2012,Garrison2014,BrookdiCintio2015a,Pap2015}. Just as in the case of the bright MW satellites, the observed kinematics of field dwarfs imply that they are hosted by low-mass halos. These halos are produced in large enough numbers in a \LCDM \ universe, that we would expect to detect significantly more dwarfs than we actually do. In the case of the field, the observational census of dwarfs usually comes from the measurement of the velocity function of galaxies \citep{Zwaan2010,Papastergis2011,Klypin2015}. Reliable measurements of the galactic VF can be carried out either using wide area surveys in the 21cm emission line of atomic hydrogen \citep[HI surveys;][]{Barnes2001,Giovanelli2005} or systematic searches of galaxies in the Local Volume \citep{Karachentsev2013}.

Given the universality of the TBTF problem among dwarf galaxies, a solution can be plausible only if it is similarly universal. Perhaps the most promising one is a ``baryonic'' solution, i.e., related to the effects that baryonic feedback processes have on low-mass halos. First, baryonic feedback can lower the number of halos that are expected to host detectable galaxies, compared to the naive expectation based on cosmological DM-only simulations \citep[e.g.,][]{Sawala2015}. The dominant process in this case is reionization feedback, which inhibits the formation of galaxies in the lowest mass halos \citep{Okamoto2008}. A second important baryonic effect involves the modification of the mass profile of a DM halo by star formation feedback. In particular, a number of studies based on hydrodynamic simulations have observed the creation of a ``core'' in the central DM density profile of low-mass halos \citep[][to name a few]{Mashchenko2008,Governato2010,Brooks2013,Madau2014,Trujillo2015,Onorbe2015}. This profile is different from the ``cuspy'' NFW profile observed for halos in DM-only simulations \citep{NFW1997}. Theoretically, core creation is attributed to repeated episodes of gas blow-out during bursts of star formation \citep{PontzenGovernato2012,Governato2012}. The energetics of star formation makes such that the efficiency of core creation differs among different galaxies, having a characteristic dependence on the stellar-to-halo mass ratio \citep{dC2014a,Chan2015} and possibly also on star formation history \citep{Onorbe2015}. Profile modification can have an important  impact on the study of the TBTF problem, because the analysis of dwarf kinematics is dependent on the assumed DM profile.

In this article we assess the TBTF problem for field dwarf galaxies, taking explicitly into account the baryonic effects described above. In Section \ref{sec:methodology} we describe in detail our methodology. Section \ref{sec:results} contains our main result, which is graphically presented in Figure \ref{fig:vhalo_vrot+gals}. In Section \ref{sec:discussion} we elaborate on the scientific implications of our result, and discuss future prospects for advancing our understanding of the TBTF problem. We conclude in Section \ref{sec:summary} by summarizing our findings.


\section{Methodology}
\label{sec:methodology}

In this article we follow closely the method developed in \citet{Pap2015}. The major novelty in this article consists of the fact that we modify certain ingredients of the P15 analysis in order to incorporate relevant effects of baryonic feedback. 

In brief, our analysis consists of the following steps: First, we determine the abundance of galaxies in the nearby universe by means of the galactic velocity function (VF). The VF of galaxies is defined as the number density of objects as a function of \Vrot, which is their rotational velocity as measured by the width of the HI emission line. We then compare the galactic VF to the corresponding distribution for halos predicted by \LCDM. The halo velocity function is defined in terms of  \Vhalo, which is the maximum rotational velocity reached by a halo's rotation curve.
Our calculation of the halo VF includes the effects of baryonic processes that can modify the abundance galaxy-hosting halos, according to the results of \citet{Sawala2015}. At this point, we avoid the common assumption that \Vrot \ $\approx$ \Vhalo, and instead use the comparison between the velocity distributions of galaxies and halos to infer the average \vrot-\vhalo \ relation that is expected in a \LCDM \ universe. This is achieved through the statistical technique of abundance matching (AM).

We then test the \vrot-\vhalo \ relation that is predicted by \LCDM, using an extensive literature sample of galaxies with spatially resolved HI kinematics. In particular, we use the rotational velocity measured for each galaxy at the outermost HI radius, \Vout$= V(R_\mathrm{out,HI})$, to constrain the mass of its host halo. This process is model-dependent, in the sense that the constraint can be different for different halo profiles. In this article, we take into account complications related to star formation feedback, by using the feedback-motivated ``DC14'' halo profile \citep{dC2014b}. In the end, each galaxy has a measured value for its HI rotational velocity and an inferred value for the mass of its host halo, allowing us to place datapoints on the \vrot-\vhalo \ plane. \LCDM \ can be considered successful only if the position of individual galaxies on the \vrot-\vhalo \ plane is consistent with the relation needed to reproduce the measured VF of galaxies.

\subsection{Comparing the number density of dwarf galaxies and low-mass halos}

\subsubsection{The velocity function of galaxies}
\label{sec:gal_velocity_function}

Currently, the most efficient way of obtaining systematic measurements of \vrot \ for large samples of galaxies is by observing their HI emission line with single-dish radio telescopes. In particular, the spectral width of a galaxy's HI profile, $W$, can be used as a measure for \vrot; this is because $W$ reflects the Doppler broadening of the HI profile caused by the line-of-sight velocity of atomic hydrogen gas as it orbits in the galactic potential. More specifically, we can define a linewidth-derived\footnotemark{} rotational velocity for a galaxy as

\begin{eqnarray}
V_\mathrm{rot,HI} = W \, / \, (2 \times \sin i) \;\;\; ,
\label{eqn:vrot}
\end{eqnarray}

\noindent
where $i$ is the inclination of the galactic HI disk to the line-of-sight.

\footnotetext{Here the HI profile width is measured at 50\% of the profile peak flux, $W_{50}$, and then corrected for the cosmological component of Doppler broadening according to the galaxy's heliocentric redshift, $W = W_{50}/(1+z_\odot)$.}

Since $W$ is one of the direct observables for a spectral line radio observation (see Fig. 1 in \citealp{Pap2015}), while $i$ is generally not, we focus here on the galactic width function (WF). The definition of the WF is directly analogous to that of the VF, but refers to the directly measured velocity width rather than the inclination-corrected rotational velocity. More specifically, the WF is calculated as

\begin{eqnarray}
n(W) = \frac{<dN_\mathrm{gal}>}{dV \, d\log_{10}W} \;\;\; .
\label{eqn:WF}
\end{eqnarray}

\noindent
In the equation above, $<dN_\mathrm{gal}>$ is the average number of galaxies with profile width values within a small logarithmic bin centered around $W$, that one can find in a volume $dV$ of our universe.

\begin{figure}
\centering
\includegraphics[scale=0.45]{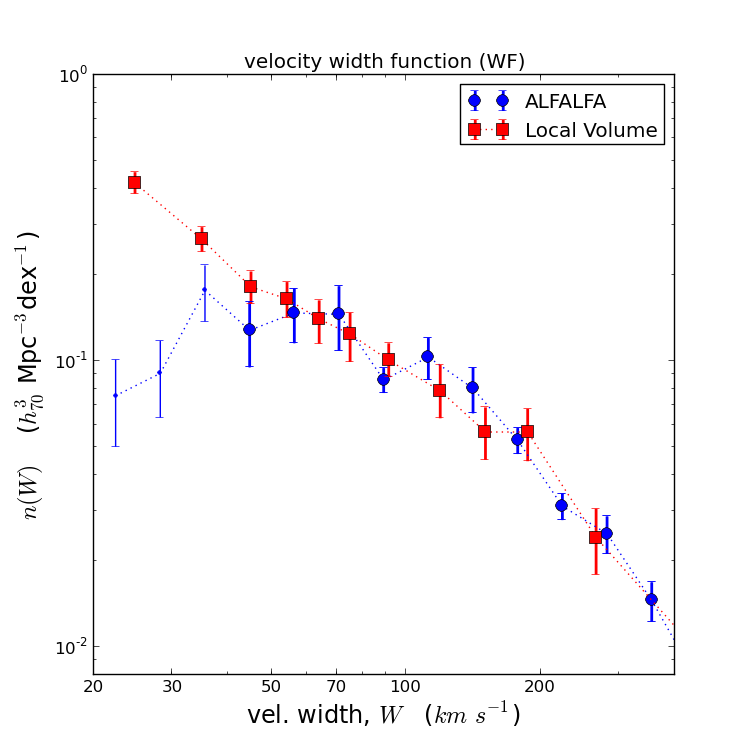}
\caption{The velocity width function (WF) of galaxies. Red squares correspond to the WF of galaxies measured in the Local Volume by \citet{Klypin2015}. The Local Volume WF is based on a nearly volume-complete optical catalog of nearby galaxies \citep{Karachentsev2013}, and includes both late-type and early-type dwarfs. Blue circles correspond to the WF measured in this work, based on the galaxy catalog of the ALFALFA HI survey \citep{Haynes2011}. The $W < 40$ \kms \ section of the ALFALFA WF (blue dots) is not used in the analysis that follows, due to possible incompleteness issues (see \S\ref{sec:gal_velocity_function} for details). In both cases, the errorbars represent the uncertainties due to counting statistics only.}
\label{fig:WF}
\end{figure}

Figure \ref{fig:WF} shows two measurements of the galactic WF in the local universe. The first corresponds to the WF measured in the Local Volume (LV; $D < 10$ Mpc) by \citet{Klypin2015}. This measurement of the WF is obtained by analyzing the catalog of nearby galaxies of \citet{Karachentsev2013}. The Karachentsev et al. catalog is optically selected, which means that it includes both late type and early type galaxies. For the majority of galaxies in the catalog (which are late-types) the velocity width is measured from their HI lineprofiles.  For gas-poor early type galaxies, the width is measured instead from their stellar kinematics (if available), or it is assigned based on their $K$-band luminosity \citep[see Fig.~1]{Klypin2015}. The LV catalog is practically volume-complete for galaxies with $W \gtrsim 40$ \kms, and so the WF is derived by a straightforward application of Eqn. \ref{eqn:WF} over this width range. The LV dataset can also probe the galactic WF to lower widths with the use of volume corrections (down to  $W \approx 25$ \kms; see Figs. 4 \& 10 in \citealp{Klypin2015}).

The second measurement of the WF is based on the dataset of the ALFALFA blind HI survey \citep{Haynes2011}. For details on the measurement methodology please refer to Appendix A in \citet{Pap2015} and references therein. Here we would like to just mention the differences between the WF measurement presented in \citet{Pap2015} and the one presented here: First, the minimum distance limit for the inclusion of a galaxy in the WF calculation has been lowered to $D_\mathrm{min} = 3$ Mpc. The measured WF will therefore include the contribution of galaxies with very low estimated distances, which typically carry a large fractional uncertainty on their distance. This will lead to a slight systematic overestimate of the WF (see \citealp[\S 4.2]{Papastergis2011}), which however is a conservative effect for the analysis performed in this article. Second, the ALFALFA WF is measured only for galaxies with $M_\mathrm{HI} \geq 10^7 \; M_\odot$ (same as in \citealp{Pap2015}). It is therefore expected that the ALFALFA measurement becomes incomplete at very low widths, due to the presence of very small late-type dwarfs with $M_\mathrm{HI} < 10^7 \; M_\odot$ and occasionally some gas-poor early type dwarfs. In this article, we use the \citet{Karachentsev2013} catalog to obtain an estimate of the ALFALFA incompleteness at low widths. We find that for $W > 40$ \kms \ the completeness of the ALFALFA WF should be very high ($\gtrsim 80\%$). As a result, we base all subsequent analysis of the ALFALFA WF on the $W > 40$ \kms \ portion, to which we apply an incompleteness correction (of up to 20\%).

As Fig. \ref{fig:WF} shows, the two measurements of the WF are fully consistent over the overlapping range of validity. However, the ALFALFA measurement has a somewhat shallower low-width slope ($\alpha \approx -1$ for the Local Volume WF versus $\alpha = -0.60 \pm 0.25$ for the ALFALFA WF). Even though the difference in slope is not statistically significant, it leads to a non-negligible discrepancy of the WF when extrapolated to very low widths (e.g., a factor of $\approx 2$ in number density at $W = 20$ \kms). This issue will later become a relevant point in the scientific discussion of this article (Sec. \ref{sec:discussion}).

We now use the two measurements shown in Fig. \ref{fig:WF} to infer the galactic VF, which is the distribution of intrinsic (i.e., de-projected) rotational velocity. The VF is denoted by $n(V_\mathrm{rot})$, and is assumed here to follow the specific analytical form of a modified Schechter function:

\begin{eqnarray}
n(V_\mathrm{rot}) = \frac{dN_\mathrm{gal}}{dV \, d\log_{10}V_\mathrm{rot}} = \ln(10) \, n^\ast \, \left(\frac{V_\mathrm{rot}}{V^\ast}\right)^\alpha \; e^{-\left(\frac{V_\mathrm{rot}}{V^\ast}\right)^\beta} \;\;\; .
\label{eqn:mod_schechter}
\end{eqnarray}

\noindent
In particular, we draw values from the distribution above and compute velocity widths according to Eqn. \ref{eqn:vrot}. In this process, we assume that galaxies are randomly oriented with respect to the line-of-sight ($\cos i$ is uniformly distributed in the $[0,1]$ interval). We then vary the set of parameters $\{n^\ast,V^\ast,\alpha,\beta\}$, until the WF that is obtained in this way matches the measured WF. The indirect statistical calculation of the VF described above is more accurate than the seemingly straightforward method based on deriving inclination-corrected rotational velocities for each individual galaxy \citep{Papastergis2011,Klypin2015}. This is because low-mass dwarfs usually have irregular shapes, and therefore optically-derived inclination values can be unreliable \citep{Obreschkow2013}.

\begin{figure*}
\centering
\includegraphics[width=\linewidth]{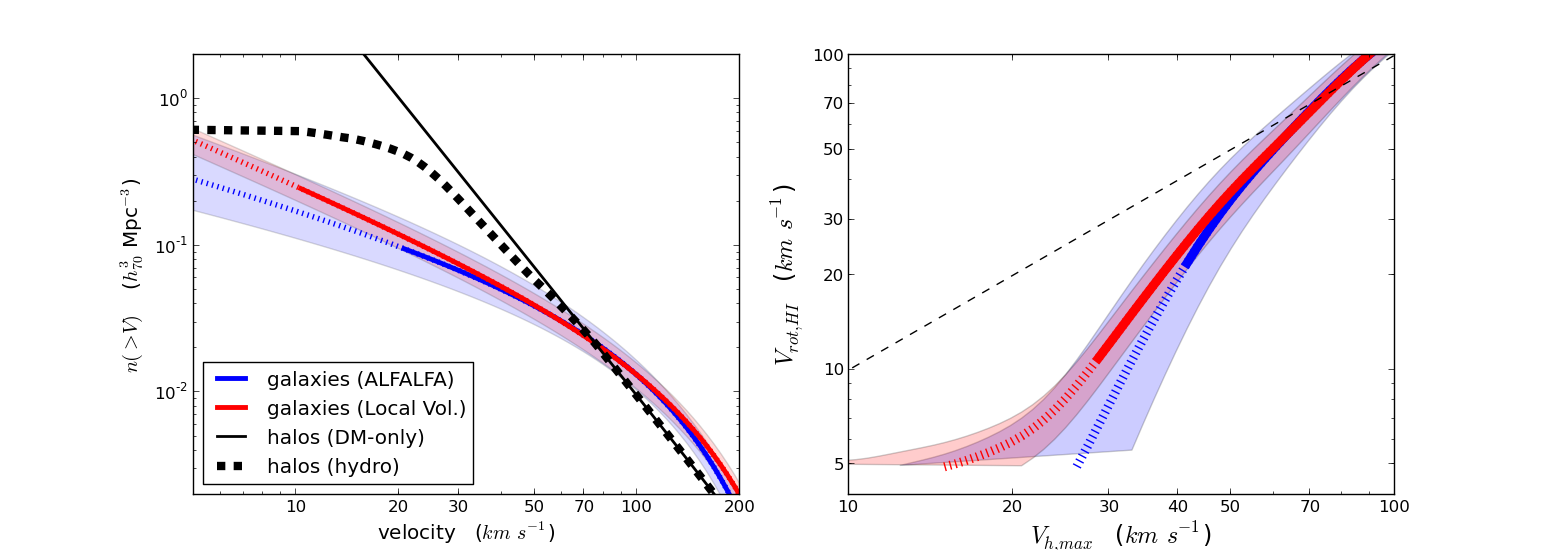}
\caption{The velocity function (VF) of galaxies and halos, and the \vrot-\vhalo \ relation according to abundance matching (AM). \textit{left panel}: The blue and red lines denote the VF of galaxies according to ALFALFA survey data (this work) and according to data for Local Volume galaxies \citep{Klypin2015}, respectively. These VFs are inferred from the corresponding direct measurements of the galactic WF (see Fig. \ref{fig:WF}). The solid portions of the lines denote the range of velocities over which each VF is constrained by data, while the dotted portions represent extrapolations. The shaded bands represent estimated errors on each observational measurement.
The black lines correspond instead to the VF of halos in a \LCDM \ universe with \textit{Planck} cosmological parameters. The thin solid line refers to the DM-only result \citep{Klypin2015}, while the thick dashed line includes ``baryonic'' effects such as baryon depletion and reionization feedback \citep{Sawala2015}. 
Note that the $x$-axis corresponds to \Vrot \ in the case of galaxies, and \Vhalo \ in the case of halos. Note also that, unlike in Fig. \ref{fig:WF}, the distributions are plotted in terms of cumulative number densities. \textit{right panel}: The blue and red solid lines are the \vrot-\vhalo \ relations that reproduce the galactic VFs measured by ALFALFA and in the Local Volume, respectively. The relations were computed by abundance matching (AM), between each of the two observational VFs and the ``baryonic'' version of the halo VF. The dotted portions of the lines are based on extrapolations of the galactic VFs. The shaded bands represent the uncertainty introduced by the observational uncertainty in the galactic VFs. The black dashed line is a reference one-to-one line.
}
\label{fig:VF+AM}
\end{figure*}

The result is shown by the colored solid lines in the left panel of Fig. \ref{fig:VF+AM}. In particular, the VF for the Local Volume has been already calculated according to the procedure above in \citet{Klypin2015}, and we adopt it here as determined in the original reference. The error band corresponds to an uncertainty of $\pm$20\%, which is the estimated systematic uncertainty of the measurement according to the original reference. The ALFALFA VF is computed in this work based on the $W > 40$ \kms \ portion of the ALFALFA measurement of the WF, and is extrapolated to lower velocities. The error band in this case reflects the datapoint scatter and the statistical errorbars of the ALFALFA WF shown in Fig. \ref{fig:WF}.

\subsubsection{The velocity function of halos in \LCDM}
\label{sec:halo_velocity_function}

The two estimates of the galactic VF obtained above can now be compared with the halo velocity function expected in a \LCDM \ universe. In particular, the left panel of Figure \ref{fig:VF+AM} shows the number density of halos as function of their maximum rotational velocity, \Vhalo. This quantity is tightly correlated with the halo virial mass \citep[e.g., see][Fig. 4]{Klypin2011}, and has the added advantage of being a definition-independent halo property. In a DM-only \LCDM \ simulation, the halo VF is approximately a power law at low velocities, with a steep exponent of $\alpha \approx -3$ \citep[e.g.,][]{Klypin2011}. A naive direct comparison between the VF of galaxies and halos reveals a huge discrepancy at low velocities: for example, there are more than an order of magnitude more halos with \Vhalo$ > 20$ \kms \ than galaxies with \Vrot$ > 20$ \kms \ \citep{Zwaan2010,Papastergis2011,Klypin2015}. 

However, there are two important effects that complicate the direct comparison described above. First, baryonic effects are expected to reduce the number density of halos that are able to host detectable galaxies. In particular, a large fraction of halos with \Vhalo$\lesssim 22$ \kms \ are expected to be dark, due to the effects of cosmic reionization in the early universe \citep[e.g.,][]{Okamoto2008}. In addition, halos with \Vhalo$\approx 25 - 70$ \kms \ are expected to be heavily baryon-depleted \citep[e.g.,][]{Papastergis2012}, and therefore are somewhat less massive than their counterparts in a DM-only simulation.

In order to capture both of these baryonic effects on the abundance of halos, we use in our analysis the ``baryonic'' VF of halos obtained by \citet{Sawala2015}. In particular, Sawala et al. use a series of hydrodynamic simulations of structure formation in the \LCDM \ context to determine the velocity distribution of halos that are able to host detectable galaxies (see the right panel of their Fig. 2). One can compare the ``baryonic'' VF of halos with the DM-only version of the distribution in the left panel of Fig. \ref{fig:VF+AM}. The two distributions start differing from one another at \Vhalo$\approx 70$ \kms, due to the effects of baryon depletion. However, the difference becomes more dramatic at \Vhalo$\lesssim 22$ \kms, where the baryonic version of the cumulative VF ``flattens out'' completely due to the effects of reionization feedback.

The second complication relates to the fact that the value of \Vrot \ --which is measured from a galaxy's HI linewidth-- is not necessarily a good measure of \Vhalo. This statement is especially true for dwarf galaxies, which often have single-peaked HI profiles suggesting that their rotation curves (RCs) do not reach a flat outer part. Therefore, a comparison of the abundance of galaxies with the abundance of halos at the same value of velocity is not necessarily a meaningful one. 
For this reason, we drop in this article the common assumption that \Vrot \ $\approx$ \Vhalo \ \citep[e.g.,][]{Zwaan2010,Papastergis2011,Klypin2015}. Instead we try to infer the relation between \Vrot \ and \Vhalo \ that reproduces the observed VF of galaxies, when applied to the ``baryonic'' VF of halos in \LCDM. This can be achieved statistically, via the technique of abundance matching (AM). In simple terms, AM consists of identifying the values of \Vrot \ and \Vhalo \ that correspond to the same cumulative number density of galaxies and halos, respectively (refer to \S 2.2 and \S 4.3 in \citealp{Pap2015} for details). Keep in mind that built into the AM methodology is the assumption that galaxies with a lower value of \Vrot \ are hosted by halos with (on average) lower values of \Vhalo. 

The \vrot-\vhalo \ relations needed to reproduce the galactic VF as measured in the Local Volume and in the ALFALFA survey are shown in the right panel of Figure \ref{fig:VF+AM}. Note how the relations drop significantly below the one-to-one line at low velocities, to compensate for the large abundance of small halos produced in \LCDM. The AM relations change behavior at \Vhalo$\lesssim 22$ \kms, reflecting the fact that very few observable galaxies are hosted by such low-mass halos.


\subsection{Constraining the host halo mass of dwarf galaxies}
\label{sec:host_halos}

\subsubsection{Galaxy sample}
\label{sec:sample}

We use the compilation of galaxies with HI kinematics drawn from the literature presented in  \citet{Pap2015}. Our sample consists of a total of 198 unique galaxies, out of which $\approx$90 can be categorized as dwarfs (\Vrot$< 50$ \kms). In particular, we have drawn 29 galaxies from the FIGGS sample \citep{Begum2008a,Begum2008b}, 18 galaxies from the THINGS sample \citep{deBlok2008,Oh2011a}, 54 galaxies from the WHISP sample \citep{Swaters2009,Swaters2011}, 5 galaxies from the sample of \citet{Trachternach2009}, 12 galaxies from the LVHIS sample \citep{Kirby2012}, 4 galaxies from the sample of \citet{Cote2000}, 30 galaxies from the sample of \citet{VerheijenSancisi2001}, 12 galaxies from the compilation of \citet{Sanders1996}, 22 galaxies from the LITTLE THINGS sample \citep{Hunter2012,Oh2015}, 11 galaxies from the SHIELD sample \citep{Cannon2011} and the dwarf galaxy LeoP \citep{Giovanelli2013,Bernstein2014}. 

A more detailed description of the sample characteristics can be found in \S3.1 of \citet{Pap2015}. As far as the galaxy sample is concerned, the largest difference with respect to our previous article is that the LITTLE THINGS and SHIELD galaxies have updated measurements of their kinematics and other observational properties \citep{Oh2015,McQuinn2015}. In addition, the kinematic analysis to be described in \S\ref{sec:kinematic_analysis} requires a new piece of information with respect to the analysis in our previous article: galactic stellar masses, $M_\ast$. Estimates for the stellar mass of our galaxies have been computed either from information in the original references, or in related articles in the literature \citep[e.g.,][]{Oh2011b}. A variety of methods was employed, depending on the availability of suitable data for each sample. For example, SHIELD galaxies have high-accuracy stellar masses based on the decomposition of the color-magnitude diagram of their resolved stellar populations \citep{McQuinn2015}. LITTLE THINGS galaxies also have high-quality stellar mass estimates based on \textit{Spitzer} photometry in the 3.6\micron \ band. On the other hand, the stellar mass estimates for FIGGS galaxies are more uncertain, since they are based on their $I$-band luminosities and $B-V$ colors (according to the mass-to-light calibration of \citealp{Bell2003}).

\subsubsection{Kinematic analysis}
\label{sec:kinematic_analysis}

In this section, we use the resolved HI kinematics data that are available for our galaxies in order to set an upper limit on the mass of their host DM halos. We follow closely the procedure that is described in \S3.2 of \citet{Pap2015}, so we encourage the reader to refer to this section for details. 

Here we give just a brief summary: Our analysis is based on the rotational velocity measured at the outermost HI radius \textit{only}, $V_\mathrm{out,HI} = V(R_\mathrm{out,HI})$. This is done not only to simplify the analysis, but also because the majority of low-mass dwarfs we analyze are drawn from the FIGGS and SHIELD samples, for which outermost point velocities and radii are the only available kinematic data. We then consider a set of halos with monotonically increasing mass, and we identify the most massive one that is compatible with the measured rotational velocity at \Rout \ (to within the 1$\sigma$ velocity uncertainty). Please refer to Figure 5 in \citet{Pap2015} for a graphical illustration of this process. In the present article, we first repeat the analysis of \citet{Pap2015}, and assume that DM halos follow the NFW density profile \citep{NFW1997}. This profile shape is motivated by DM-only simulations of structure formation. For each halo, the concentration, $c$, is fixed according to the mean mass-concentration relation in a \textit{Planck} cosmology \citep[Table 3]{DuttonMaccio2014}. In Appendix \ref{sec:appendix_b} we also consider the case of scatter in halo concentrations about the mean mass-concentration relation.

We then repeat our kinematic analysis using a halo profile that is motivated by hydrodynamic simulations of galaxy formation, namely the ``DC14'' profile \citep{dC2014b}. This is done in order to assess to what extent profile modification by baryonic feedback can provide a solution to the TBTF problem. More specifically, feedback effects can create a central ``core'' in the DM density profiles of halos. Halos with cored profiles display lower rotational velocities in their inner regions compared to their ``cuspy'' NFW counterparts. If dwarf galaxies in nature have indeed cored inner profiles \citep[e.g.,][]{Oh2011b}, an NFW analysis of their observed galactic kinematics could lead to unrealistically small inferred masses for their host halos.

\citet{dC2014a} find that the efficiency of core creation is primarily determined by the stellar-to-halo mass ratio of a galaxy, $M_\ast/M_\mathrm{h}$. In order to capture this effect with a simple analytical prescription, \citet{dC2014b} first parametrize the DM density profile with a double power-law functional form,



\begin{eqnarray}
\rho(r) = \frac{\rho_s}{\left( \frac{r}{r_s} \right)^\gamma \left[ 1 + \left( \frac{r}{r_s} \right)^\alpha \right]^{(\beta-\gamma)/\alpha}}  \;\;\; .
\label{eqn:double_powerlaw}
\end{eqnarray}

\noindent
In the equation above, $r_s$ and $\rho_s$ are the scale radius and scale density of the profile; as in the NFW analysis, these values are set based on the mean mass-concentration relation of halos in a \textit{Planck} cosmology. The parameters $\alpha$ and $\beta$ correspond instead to the power law exponents of the density profile in the inner and outer parts of the halo, while $\gamma$ controls the transition sharpness. The NFW profile corresponds to the parameter values $\alpha =1$, $\beta=3$, $\gamma =1$. In the case of the DC14 profile, the three shape parameters $\alpha,\beta,\gamma$ are not fixed, but vary according to analytic functions of $M_\ast/M_h$. These functions are calibrated so as to reproduce the density profiles of halos in a set of hydrodynamic simulations (Eqn. 3 in \citealp{dC2014b}). In addition, DC14 profiles may include a baryonic modification of their concentration parameter \citep[Eqn. 6 in][]{dC2014b}. This modification concerns only halos hosting galaxies with high stellar-to-halo mass ratios ($\log(M_\ast/M_h) > -2$), and accounts for the effect of adiabatic contraction on the halo profile. A very useful guide for the practical implementation of DC14 profiles can be found in the Appendix of \citet{dC2014b}.

As far as dwarf galaxies are concerned, the main feature of the DC14 profile is that galaxies with stellar-to-halo mass values in the range of $\log(M_\ast/M_h) \approx$ -3 to -2 have a cored inner density profile. If we consider for example a host halo with \Vhalo$ \approx 40$ \kms, then a core is expected to form if the hosted dwarf has a stellar mass in the range $M_\ast \approx 10^{7} \; \mathrm{to} \;  10^{8} \; M_\odot$. On the other hand, galaxies with significantly lower stellar-to-halo mass ratios, $\log(M_\ast/M_h) \lesssim -4$, are not efficient at modifying their underlying DM density distribution, because there is not enough supernova energy available. As a result, if the same example halo were hosting a dwarf with $M_\ast \lesssim 10^{6} \; M_\odot$, then its DM profile would be expected to follow the cuspy NFW form.

\begin{figure}
\centering
\includegraphics[width=\columnwidth]{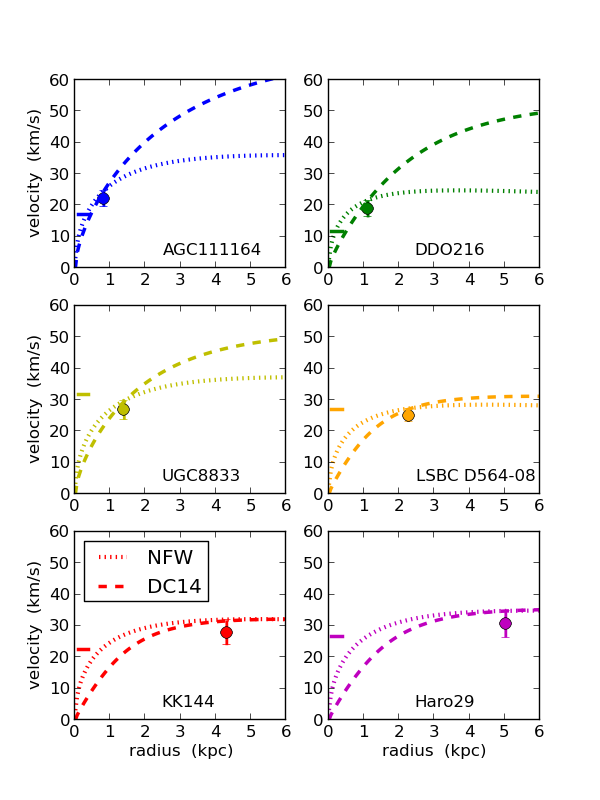}
\caption{
Datapoints with errorbars represent the rotational velocity measured at the outermost HI radius, \Vout=$V(R_\mathrm{out,HI})$, for six example galaxies. 
The galaxies are ordered from top left to bottom right in order of increasing \Rout. In each panel, the dotted and dashed lines correspond to the RCs of the most massive halo that is compatible with the observed datapoint within 1$\sigma$, assuming an NFW and DC14 profile respectively. The horizontal mark denotes the value of the linewidth-derived \Vrot \ for each galaxy.
}
\label{fig:galaxy_panel}
\end{figure}

Figure \ref{fig:galaxy_panel} shows the results of the kinematic analysis described above for six example galaxies, using both an NFW and a DC14 profile\footnotemark{}. All six galaxies are extreme dwarfs that suffer from the TBTF problem in the NFW analysis (i.e., they are located clearly to the left of the AM relation in Fig. 6 of \citealp{Pap2015}). It is important to keep in mind that the halo RCs plotted in Fig. \ref{fig:galaxy_panel} correspond in reality to upper limits; this is because we do not subtract the velocity contribution of stars and gas from the value of \Vout \ when constraining the host halo mass.

\footnotetext{Unlike in the case of mean concentration NFW halos --where the rotational velocity at any radius increases monotonically with increasing halo mass-- DC14 halos can display a non-monotonic dependence of their rotational velocity with halo mass at small radii. We are aware of this complication when fitting DC14 profiles to our sample of galaxies.}

The galaxies are placed in the figure in order of increasing outermost HI radius, \Rout. 
This order serves to illustrate a trend that has very important implications for assessing whether core creation can provide a solution to the TBTF problem.
In particular, galaxies with small HI radii, \Rout$\lesssim 1.5$ kpc, place very different constraints on the mass of their host halo depending on whether a cuspy NFW or a cored DC14 profile is assumed. For example, galaxy DDO216, which has \Rout$= 1.12$ kpc, has a ratio of $V_\mathrm{h,DC14} \, / \, V_\mathrm{h,NFW} = 2.11$ (see top right panel of Fig. \ref{fig:galaxy_panel}). On the other hand, galaxies with larger HI radii, \Rout$\gtrsim 2$ kpc, place approximately the same constraint on \vhalo \ for both profile shapes. An example is galaxy Haro29, which has \Rout$= 5.03$ kpc and $V_\mathrm{h,DC14}/V_\mathrm{h,NFW} = 1.01$ (see bottom right panel of Fig. \ref{fig:galaxy_panel}).

The trend above has a simple and intuitive explanation: Core creation due to feedback is limited to the radial extent over which star-formation takes place in dwarf galaxies, typically $R \lesssim 1$ kpc \citep[see, e.g., Fig. 4 in][]{PontzenGovernato2012}. If the kinematics of a galaxy can only be measured over this kind of length-scale, then the existence or not of a central core has a profound effect on the inferred properties of the host halo. 
Conversely, galaxies with more extended HI disks, \Rout$\gtrsim 2$ kpc, probe the region of the profile that is not heavily affected by stellar feedback. As a result, the constraint that they can place on the mass of their host halo is roughly independent of the shape of the DM density profile in the inner region.


\section{Results}
\label{sec:results}

\begin{figure*}
\centering
\includegraphics[scale=0.42]{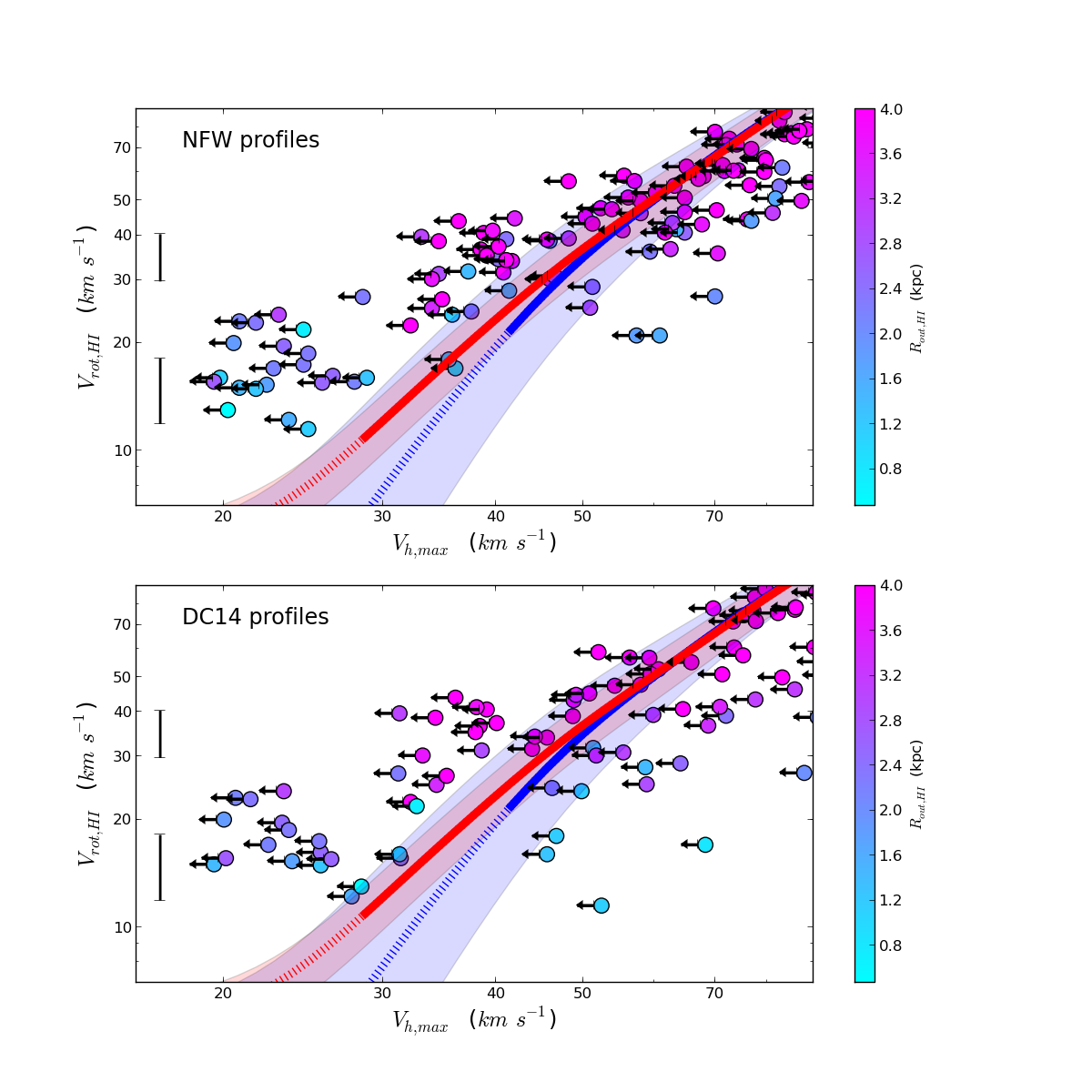}
\caption{\textit{top panel}: Datapoints represent the position of our sample of galaxies in the \Vhalo\ - \Vrot \ plane. The $y$-axis position corresponds to the rotational velocity derived from the inclination-corrected HI linewidth, \Vrot. The vertical errorbars represent typical $y$-axis errors for galaxies with \Vrot=15 \kms \ and 35 \kms. The $x$-axis position of each galaxy corresponds instead to the upper limit on the \Vhalo \ of its host halo, inferred from the galactic kinematics measured at the outermost HI radius and assuming an NFW density profile (refer to \S\ref{sec:kinematic_analysis}).
The color-coding represents the radius of the outermost velocity measurement, \Rout. The blue and red shaded regions represent the AM relations between \Vhalo \ and \Vrot \ that reproduce the galactic velocity function measured in the ALFALFA survey (this work) and within the Local Volume \citep{Klypin2015}. 
\textit{bottom panel}: Same as in the top panel, but galaxies are placed on the $x$-axis assuming their host halos follow the DC14 density profile. Please refer to Sec. \ref{sec:results} for the scientific interpretation of this figure.}
\label{fig:vhalo_vrot+gals}
\end{figure*}

Figure \ref{fig:vhalo_vrot+gals} shows the placement of galaxies in the \Vrot \ -- \Vhalo \ plane, according to the NFW and DC14 analysis (top and bottom panel, respectively). In both panels, the $y$-axis is \Vrot, which corresponds to the inclination-corrected HI linewidth of galaxies (Eqn. \ref{eqn:vrot}; see also Fig. 1 in \citealp{Pap2015}). Typical $y$-axis errorbars for galaxies with \Vrot= 15 \kms \ and 35 \kms \ are also plotted for reference. These errorbars are derived from Eqn. \ref{eqn:vrot} assuming typical values for the observational error on the measured HI linewidth, $\sigma_{W_{50}} = 4$ \kms, and on the galactic inclination, $\sigma_i = 10^\circ$. Note that the error on \Vrot \ is different for each object, because it strongly depends on the inclination of each individual galaxy. The plotted errorbars refer to an indicative inclination of $i = 57^\circ$, which is the average value for randomly oriented objects. A more detailed explanation of the impact of observational errors on the placement of galaxies in the \vrot-\vhalo \ plane can be found at the end of \S3.3 in \citet{Pap2015}. 

The $x$-axis corresponds instead to the upper limit on \Vhalo \ for each galaxy's host halo, inferred from the analysis of the resolved HI kinematics of each object (refer to \S\ref{sec:kinematic_analysis}). Keep in mind that, unlike \Vrot, \Vhalo \ is not an observational property of a galaxy. It is rather a model-dependent quantity, in the sense that the inferred value depends on the assumed DM profile shape. In both panels, the datapoints are color-coded according to the value of the galactic outermost HI radius, \Rout. As explained in \S\ref{sec:kinematic_analysis}, galaxies with \Rout$\lesssim 1.5$ kpc can have a large difference between the upper limit on \Vhalo \ obtained using an NFW or a DC14 profile. As a result, cyan shaded datapoints can move a lot in the $x$-axis direction between the two panels. On the other hand, galaxies with larger outermost radii, \Rout$ \gtrsim 2$ kpc, can place approximately the same constraint on their host halo mass irrespective of the assumed profile. Therefore, magenta shaded datapoints are expected to keep a similar $x$-axis position in both panels.

The blue and red shaded regions in Figure \ref{fig:vhalo_vrot+gals} correspond to the relations between \Vhalo \ and \Vrot \ that reproduce the ALFALFA and Local Volume measurements of the galactic VF, respectively (see Sec. \ref{sec:gal_velocity_function}). In the top panel of Fig. \ref{fig:vhalo_vrot+gals}, datapoints with \Vrot$\lesssim 25 - 30$ \kms \ are incompatible with the average relations needed to reproduce the observed velocity function of galaxies. This constitutes the well-established TBTF problem for field galaxies \citep{Ferrero2012,Garrison2014,Pap2015,BrookdiCintio2015a}. Put simply, the problem consists of the fact that the kinematics of the smallest dwarfs in our sample indicate that they are hosted by NFW halos with \Vhalo$= 20 -30$ \kms; however, halos of this size are too abundant in a \LCDM \ universe to explain the low number density that is observed for such dwarfs.

The situation is different in the DC14 analysis shown in the bottom panel of Figure \ref{fig:vhalo_vrot+gals}. The datapoints display increased scatter, due to several objects moving towards larger values of \Vhalo. Objects that are compatible with the AM relations represent now a sizable minority of all galaxies with \Vrot$< 25$ \kms. However, it is crucial to notice that there is a clear segregation between the datapoints that are compatible with the AM relations and those that are not: the former are mostly cyan shaded, indicating small values of \Rout.
On the other hand, the majority of our galaxies remain incompatible with the AM relations, including all objects with relatively large values of \Rout \ (blue and magenta shaded datapoints). Since these latter type of objects can place firmer constraints on the mass for their host halos, we conclude that adopting DC14 profiles in our kinematic analysis has not resolved the TBTF problem. 

Let us note that the results presented in Fig. \ref{fig:vhalo_vrot+gals} are robust against a number of theoretical complications. For example, including scatter in the AM procedure described in \S\ref{sec:halo_velocity_function} will lead to an average relation that is displaced to slightly higher values of \Vhalo. As a result, the zero-scatter case considered in this work represents a conservative choice for assessing the TBTF problem. Furthermore, we argue in Appendix \ref{sec:appendix_b} that including scatter in halo concentration as part of our kinematic analysis (\S\ref{sec:kinematic_analysis}) does not change our results substantially. Last and most importantly, the result shown in the bottom panel of Fig. \ref{fig:vhalo_vrot+gals} is not specific to the particular parametrization of the DC14 profile. In fact, different hydrodynamic simulations can lead to different calibrations of the parameters of Eqn. \ref{eqn:double_powerlaw} on the stellar-to-halo mass ratio \citep[e.g.,][]{Chan2015}. The fact that most of our extreme dwarf galaxies have kinematic measurements at relatively large radii makes our results insensitive to the detailed shape of the inner halo profile. In general, any ``core'' model \citep[e.g.,][]{delPopolo2016,Elbert2015,Fry2015,Ogiya2015} can only impact the results of this article if it can strongly affect the kinematics out to $R \approx 2 - 3$ kpc.   


\section{Discussion and future prospects}
\label{sec:discussion}

Baryonic effects are certainly important in the context of the field TBTF problem, but the result of Figure \ref{fig:vhalo_vrot+gals} suggests  that their effect is not large enough to fully resolve the issue. This conclusion seems to be at odds with the conclusions of a number of literature works on the topic. For example, \citet{Sawala2015} claim that baryonic effects on the number density of halos alone are enough to reconcile \LCDM \ expectations with observations. In this article, we have incorporated in our analysis their ``baryonic'' modification to the velocity function of halos, but we were nonetheless  unable to account for the discrepancy between theory and observations (top panel of Fig. \ref{fig:vhalo_vrot+gals}). This is because the field TBTF problem refers to halos with \Vhalo$\approx 25 - 45$ \kms, which are too massive to be affected by reionization feedback (see also \S4.3 in \citealp{Pap2015}). In fact, it is the effect of baryon depletion on the halo VF --rather than the effect of reionization-- that results in a small alleviation of the field TBTF problem. The same conclusion regarding the issue was reached by other works in the literature, for example \citet[see the left panel of their Fig. 2]{BrookdiCintio2015a} and \citet{Pawlowski2015}. A close look at Fig. 4 of \citet{Sawala2015} can also confirm our conclusion regarding reionization feedback: notice how the ``bend'' of the AM relation induced by reionization plays no role in explaining the properties of the plotted dwarf irregulars. At the same time, we would like to point out that all of the aforementioned literature articles use AM relations between stellar and halo mass to assess the TBTF problem (i.e., ``$M_\ast$-$M_\mathrm{h}$'' relations). This fact complicates the comparison between the results of these previous studies and the results obtained in this work (please refer to Appendix \ref{sec:appendix_a}).

The impact of halo profile modification on the TBTF problem is even more difficult to assess. This is evident when comparing Fig. \ref{fig:vhalo_vrot+gals} in this article to Fig. 2 in \citet{BrookdiCintio2015a}. Given that the kinematic analysis method is the same in both works, it is surprising that the two results are in such stark contrast. More specifically, both works agree that there is a TBTF problem for field galaxies when their kinematics are analyzed based on NFW density profiles. However, \citet{BrookdiCintio2015a} find that the TBTF problem is entirely solved when DC14 profiles are used instead. The difference in the two results can be understood in terms of the different dwarf samples employed by the two studies. \citet{BrookdiCintio2015a} use a sample of isolated dwarf galaxies in the Local Group ($D \lesssim 1.5$ Mpc), which have stellar kinematic measurements based on optical spectroscopy \citep{Kirby2014}. For most objects, the stellar kinematics probe their velocity field only out to \rout$ \lesssim 1 - 1.5$ kpc. As discussed in \S\ref{sec:kinematic_analysis}, objects with measured kinematics over this range of radii have very small constraining power on the mass of their host halo when a cored profile is assumed. On the other hand, our sample consists of galaxies in the Local Volume ($D \lesssim 10$ Mpc), that have interferometric observations in the HI line. This allows us to probe the kinematics for the majority of our objects out to radii $> 2$ kpc. 

The discussion above highlights the importance of having a \textit{representative} sample of dwarf galaxies with kinematics measured at a galactocentric \textit{radius} that is \textit{as large as possible}. At present, neither the sample of isolated Local Group dwarfs of \citet{Kirby2014} nor the sample of dwarfs compiled in this article satisfy both conditions. In particular, the former set of dwarfs has stellar kinematics measurements only, which tend to be limited in radial extent. On the other hand, our sample has more spatially extended HI kinematic measurements, but it consists of galaxies that have been specifically targeted for HI interferometric observations. This means that our sample could be over-representing dwarfs with extended HI disks\footnotemark{}. 
The selection of our sample, therefore, allows for the possibility that our conclusions drawn from the bottom panel of Fig. \ref{fig:vhalo_vrot+gals} are not valid. This would be the case, for example, if the bulk of objects in an unbiased sample of field dwarfs turned out to have relatively small HI disks. One could then argue that the kinematics of our sample of field dwarfs are not representative of the average dwarf population, but objects hosted by low mass halos are favored in our sample due to some selection effect. Significant progress in this area will be achieved with the next-generation HI interferometric surveys, such as the medium-deep and shallow surveys with the WSRT radio array \citep{Verheijen2008} and the WALLABY survey with the ASKAP radio array \citep{Duffy2012}. Such surveys will measure the HI kinematics for samples of nearby dwarfs that are blindly detected within the survey area. 

\footnotetext{In general, targeted interferometric campaigns select objects that are bright in the HI line. Due to the $M_\mathrm{HI} \propto R_\mathrm{HI}^2$ correlation \citep{VerheijenSancisi2001}, these galaxies also tend to have fairly extended HI disks. An exception is the SHIELD sample, which is selected to contain galaxies with $M_\mathrm{HI} < 10^{7.7} \; M_\odot$. The SHIELD selection favors therefore galaxies with relatively small HI disks.}

\begin{figure*}
\centering
\includegraphics[width=\linewidth]{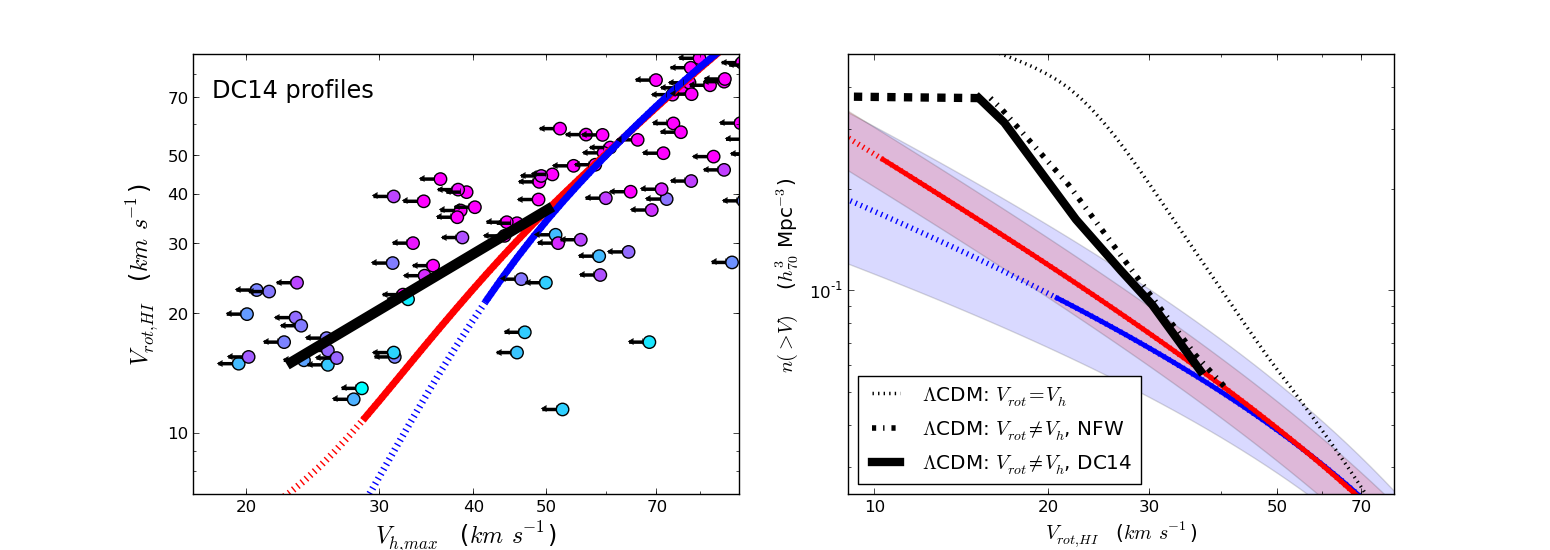}
\caption{\textit{left panel}: Similar to the bottom panel of Figure \ref{fig:vhalo_vrot+gals} (error bands omitted for clarity). The thick black line corresponds to our by-eye estimate of the average \Vrot-\Vhalo \ relation implied by the dwarf data, when analyzed with DC14 profiles. Dwarfs with low values of \Rout \ (cyan datapoints) are given low weight in our estimate (see \S\ref{sec:kinematic_analysis}). The relation stops at \Vhalo$\approx 22$ \kms, because we assume that smaller halos do not host galaxies due to cosmic reionization \citep{Sawala2015}. \textit{right panel}: Similar to the left panel of Figure \ref{fig:VF+AM}. The thick black line denotes the cumulative VF of galaxies that results from the average \vrot-\vhalo \ relation implied by the dwarfs (i.e., according to the black line in the left panel of this figure). It should be interpreted as a lower limit on the galactic VF that is expected in \LCDM, after accounting for baryonic effects (see Sec. \ref{sec:discussion}). 
The dashed portion of the line reflects the abrupt halt in the growth of the galaxy count, that is imposed to mimic the effect of cosmic reionization. For reference, we also show the \LCDM \ expectation for the galactic VF when no profile modification by baryons is assumed (black dot-dashed line), and when \Vrot$=$\Vhalo \ is assumed in addition (dotted black line, same as thick dotted line in the left panel of Fig. \ref{fig:VF+AM}).}
\label{fig:theory_vs_observations}
\end{figure*}

Another central aspect regarding the result of this article has to do with the magnitude of the discrepancy between theory and observations that is implied by Fig. \ref{fig:vhalo_vrot+gals}. In particular, both panels of the figure point to an over prediction of the number of dwarf galaxies in the \LCDM \ model; however, it is not straightforward to quantify the excess based on these plots. In Figure \ref{fig:theory_vs_observations} we make a quantitative estimate of the discrepancy, based on the position of dwarf galaxies in the \vrot-\vhalo \ plane. In the left panel of Figure \ref{fig:theory_vs_observations}, we make a by-eye estimate of the average \vrot-\vhalo \ relation at low velocities that is implied by the datapoints in the DC14 analysis. We give less weight to the position of galaxies with small values of \Rout, because of their limited constraining power on \Vhalo \ (see discussion in \S\ref{sec:kinematic_analysis} and Sec. \ref{sec:results}). We then perform ``inverse AM'' of the average \vrot-\vhalo \ relation implied by the data with the ``baryonic'' version of the halo VF \citep{Sawala2015}. This allows us to infer the cumulative number density of galaxies as a function of \Vrot \ that is expected in \LCDM, after all baryonic effects have been taken into account. Note that, since the dwarf datapoints correspond to upper limits in \Vhalo, this procedure can only be used to set a \textit{lower limit} on the number of galaxies expected in \LCDM.

In the right panel of Fig. \ref{fig:theory_vs_observations} we compare the \LCDM \ expectation for the galactic VF to the distributions measured by ALFALFA (this work) and in the Local Volume \citep{Klypin2015}. Indeed, the theoretical expectation is not compatible with the observational measurements at low velocities. However, the discrepancy is not very large: If one considers the upper error envelope of the observations, then \LCDM \ predicts a factor of $\approx$ 1.8 more dwarf galaxies with \Vrot$> 15$ \kms \ than observed. If one considers instead the central fits for the two observational VFs, then the \LCDM \ excess is a factor of $\approx 2.3 - 3$.

Given the relatively small magnitude of the difference between theory and observations, it is of paramount importance to obtain a 
a measurement of the galactic VF that is as accurate as possible. At present, the accuracy of the Local Volume measurement is limited by cosmic variance and by the incompleteness of the LV catalog at very low luminosities (see Sec. 4 in \citealp{Klypin2015}). Progress on this front can be achieved by further improving on the work of \citet{Karachentsev2013}, which would lead to ever more complete catalogs of nearby faint dwarfs. On the HI survey front, progress can come from the next generation of blind surveys, to be performed both by interferometric facilities (e.g., WSRT, ASKAP, SKA) or by single-dish radio telescopes (e.g., FAST, Arecibo AO40). These future surveys will push the detection limits to lower HI masses and larger distances, reducing the two principal sources of bias for the HI measurement of the VF (incompleteness and distance uncertainties; see Sec. 5 in \citealp{Zwaan2010} and Sec. 4 in \citealp{Papastergis2011}).

Lastly, note that several theoretical works have attempted to directly model the galactic VF expected in \LCDM, based on simulation results \citep[to name a few]{Trujillo2011,Sawala2013,Obreschkow2013,BrookdiCintio2015b,BrookShankar2015,Yaryura2016}. This approach has the advantage of partially sidestepping the AM procedure involved in the test of Fig. \ref{fig:vhalo_vrot+gals}. However, keep in mind that most of the complications pointed out in this work are still present for this type of modeling. For example, results depend critically on the galactic radii over which kinematics are assumed to contribute to the HI profile width (e.g., \citealp{BrookdiCintio2015b}, Fig. 3; see also \citealp{BrookShankar2015}). 
In addition, AM-based results are still typically used in these works, to assign various baryonic properties to halos (e.g. stellar masses, which also determine indirectly the kinematic radii). 
Results will become more robust once baryonic properties and HI lineprofiles are self-consistently extracted from the hydrodynamic simulations themselves, rather than modelled \citep{BrooksPapastergis2016}. Eventually, hydrodynamic simulations with the necessary resolution and complexity to model the HI distribution of small field dwarfs will be run over cosmological volumes. Then it will be possible to produce theoretical expectations for the galactic VFs by direct application of Eqn. \ref{eqn:WF}. At that point, the theoretical expectation for the galactic VF will be solely dependent on the assumed cosmological model and the implementation of baryonic physics.



\section{Summary}
\label{sec:summary}

In this article we assess whether the TBTF problem is present for field dwarf galaxies, once baryonic effects on the abundance and profile shape of low-mass halos are taken into account. We first determine the velocity function (VF) of dwarf galaxies, based on the dataset of the ALFALFA blind HI survey (this work) and also on a catalog of Local Volume galaxies \citep{Klypin2015}. The galactic VF measures the number density of objects as a function of \Vrot, which is the rotational velocity derived from the width of the galactic HI lineprofile (\S\ref{sec:gal_velocity_function}). We then calculate the number density of halos expected to host galaxies in \LCDM, as a function of their peak circular velocity, \Vhalo \ (\S\ref{sec:halo_velocity_function}). Our calculation includes the effects of baryon depletion and cosmic reionization on the number density of galaxy-bearing halos, according to the hydrodynamic simulations of \citet{Sawala2015}. The galactic and halo VFs are not directly comparable, as the HI disks of dwarfs do not usually extend out to the radius where the halo rotational velocity achieves its maximum value. We use instead the technique of abundance matching (AM) to statistically match the two velocity distributions (\S\ref{sec:halo_velocity_function}). 
This allows us to infer the average relation between \Vrot \ and \Vhalo \ that reproduces the observed VF of dwarf galaxies in a \LCDM \ universe.

We then assemble a comprehensive sample of galaxies from the literature that have measurements of their spatially resolved kinematics in the HI line. This sample, which contains $\approx$90 dwarfs, is used to test the \vrot-\vhalo \ relation that is expected in \LCDM. In particular, for each galaxy we use the velocity at the outermost HI radius, \Vout$=V($\Rout$)$, to identify the most massive halo that is compatible with the observed galactic kinematics (\S\ref{sec:kinematic_analysis}). This process allows us to set an upper limit on \Vhalo \ for each dwarf's host halo, and allows us to place individual objects on the \vrot-\vhalo \ plane (Fig. \ref{fig:vhalo_vrot+gals}). The constraint on \Vhalo \ depends on the density profile that is assumed for the host halo; in this article, we use both a DM-only motivated NFW profile, as well as a feedback motivated DC14 profile \citep{dC2014b}.
The NFW analysis reveals that the host halos of dwarfs with \Vrot$\lesssim 25 - 30$ \kms \ are smaller than what expected based on the AM relation (Sec. \ref{sec:results}). This is the TBTF problem for field dwarfs, which is already well established in the literature \citep{Ferrero2012,Garrison2014,Pap2015,BrookdiCintio2015a}.

Most importantly, we find that the TBTF problem is present in our DC14 analysis, as well. This finding is in direct contrast with the results of previous work on the same topic, which finds that core creation provides a solution to the TBTF problem for both satellite and isolated dwarfs (compare Fig. \ref{fig:vhalo_vrot+gals} in this article with Fig. 2 in \citealp{BrookdiCintio2015a}). 
We attribute this tension to the fact that core creation can explain the low rotational velocities observed for dwarf galaxies only in the innermost $\sim$1.5 kpc (see \S\ref{sec:kinematic_analysis} and Fig. \ref{fig:galaxy_panel}). \citet{BrookdiCintio2015a} used in their study a sample of isolated Local Group dwarfs \citep{Kirby2014}, with measurements of their stellar kinematics that probe precisely this range of radii. On the other hand, our sample of field dwarfs has HI kinematic measurements that extend in many cases out to \Rout \ $\approx 2 - 3$ kpc (and even beyond for a few objects), where the kinematic impact of the core is small. The fact that the HI velocities measured at these larger radii are still observed to be lower than expected based on the AM relations, leads us to the conclusion that core creation does not provide a full solution to the TBTF problem.

However, there is still the possibility that our conclusion above is not correct. For example, the analysis in this work assumes that there are no serious observational biases or data quality issues \citep[e.g.,][]{Oman2016,Read2016}, and that HI kinematics reflect the true enclosed dynamical mass. Most importantly, our sample of dwarfs with resolved kinematics consists of objects specifically targeted for HI interferometric observations. As a result, there is no guarantee that they are representative of the general population of field dwarfs. 

In any case, we find that the difference between the \LCDM \ expectation regarding the abundance of dwarf galaxies and observations is not very large: given the current observational uncertainties, it could be as small as a factor of $\approx 1.8$ for galaxies with \Vrot$> 15$ \kms.
The relatively small magnitude of the discrepancy between the observed number density of dwarfs and the theoretical expectation in \LCDM \ means that obtaining a more accurate measurement of the galactic VF in the future will be of great scientific importance. Progress on the observational front can be achieved with upcoming large-area HI surveys with interferometric arrays (such as WSRT and ASKAP, and eventually SKA). These surveys will provide blindly selected samples of dwarfs with spatially resolved velocity information. In addition, upcoming HI surveys with both interferometers and single-dish radiotelescopes will push the HI sensitivity limits relative to ALFALFA, delivering better measurements of the velocity function at the low end. 


\begin{acknowledgements}

The authors acknowledge the work of the entire ALFALFA
collaboration team in observing, flagging, and extracting the catalog of HI galaxies used in this work. The authors would also like to thank John Cannon for sharing insights on SHIELD data. E.P. is supported by a NOVA postdoctoral fellowship at the Kapteyn Institute. The ALFALFA team at Cornell is supported by U.S. NSF grant AST-1107390 to M.P. Haynes and R. Giovanelli, and by grants from the Brinson Foundation.

\end{acknowledgements}

\appendix

\section{TBTF assessment based on the $M_\ast$-$M_h$ relation}
\label{sec:appendix_a}

\begin{figure*}
\centering
\includegraphics[scale=0.52]{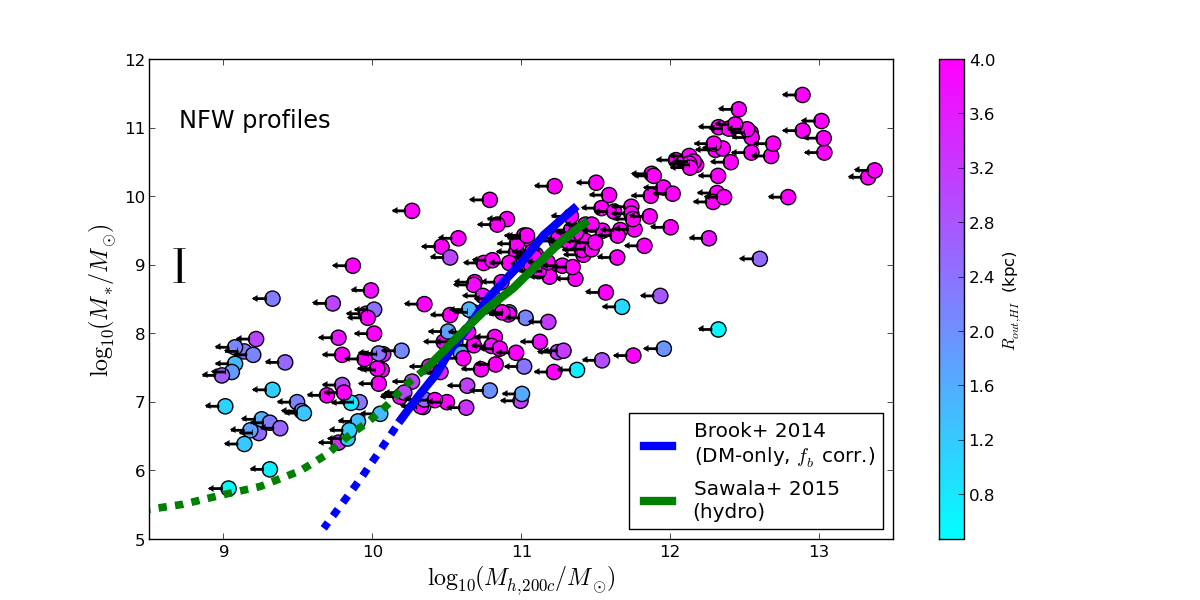}
\caption{ Datapoints represent the position of galaxies in our sample in the \mstar-\mhalo \ plane. The $x$-axis position of each galaxy corresponds to the virial mass of the most massive NFW halo that is compatible with the galactic kinematics measured at the outermost HI point (see \S\ref{sec:kinematic_analysis}). The $y$-axis position corresponds to the estimated stellar mass of each object (see \S\ref{sec:sample}). The datapoints are color-coded based on their outermost HI radius, \Rout. The blue and green solid lines represent the AM relations between \mstar \ and \mhalo \ obtained by \citet{Brook2014} and \citet{Sawala2015}, respectively. The dashed portions of the lines are extrapolations of the derived relations. A reference errorbar is also plotted, representing the typical uncertainty in the stellar mass estimates for our galaxies ($\approx 0.25$ dex). Please refer to Appendix \ref{sec:appendix_a} for a more detailed description of this figure, and for its scientific interpretation. }
\label{fig:mhalo_mstar+gals}
\end{figure*}

Figure \ref{fig:mhalo_mstar+gals} shows the placement of our sample of galaxies with resolved HI kinematics in the \mstar-\mhalo \ plane. Galaxies are placed on the $x$-axis of the figure according to the kinematic analysis described in \S\ref{sec:kinematic_analysis}, and in particular the one assuming an NFW halo profile. In this figure, the most massive compatible halo is characterized by its virial mass, \mhalo, rather than its maximum circular velocity, \Vhalo. The virial mass is defined at an overdensity of 200 with respect to the critical density, i.e. \mhalo$=$\Mhalo. The $y$-axis position of each galaxy corresponds instead to its estimated stellar mass, \mstar \ (see \S\ref{sec:sample}).

Figure \ref{fig:mhalo_mstar+gals} also shows the average \mstar-\mhalo \ relations expected in \LCDM \ according to abundance matching (AM). These AM relations are obtained by matching the mass function of halos in \LCDM, $n_\mathrm{h}(>M_\mathrm{h})$, with the stellar mass function (SMF) of galaxies, $n_\mathrm{gal}(>M_\ast)$. The first AM relation shown is the one determined by \citet{Brook2014} based on the stellar masses of galaxies in the Local Group \citep{McConnachie2012}. The \mstar-\mhalo \ relation of Brook et al. is reasonably well constrained by current observations down to \mstar$\approx 10^{6.5} \; M_\odot$, and is extrapolated below this stellar mass value. Note that the relation of Brook et al. has been shifted to lower halo masses by a factor of $(1-f_b)$, where $f_b$ is the cosmic baryon fraction. This has been done to mimic the effect of baryon depletion on the number density of low-mass halos. 

The second relation is that derived by \citet{Sawala2015}, by including baryonic effects on the \mstar-\mhalo \ relation of \citet{Moster2013}. The Moster et al. AM relation is in turn based on the SMF of galaxies measured in the Sloan Digital Sky Survey (SDSS) by \citet{LiWhite2009} and \citet{Baldry2008}. The SDSS data can constrain the SMF down to $M_\ast \approx 10^{7.5} \; M_\odot$, below which an extrapolation is used. Note that the baryonic correction of Sawala et al. reflects the results of a set of hydrodynamic simulations, and so includes the effects of both baryon depletion and reionization feedback on the number density of galaxy-bearing halos.

The purpose of Figure \ref{fig:mhalo_mstar+gals} is to provide a way of directly comparing our work with the results of relevant studies in the literature (e.g., \citealp[Fig. 12]{Garrison2014}; \citealp[top panel of Fig. 4]{Sawala2015}; \citealp[left panel of Fig. 2]{BrookdiCintio2015a}). The first thing to notice is that there is no clear evidence for the TBTF problem in the \mstar-\mhalo \ plane. Low-mass galaxies are positioned on either side of the AM relations in roughly equal proportions. This is in contrast with what seen in the corresponding analysis based on the velocity function, shown in the top panel of Fig. \ref{fig:vhalo_vrot+gals}. Instead, Fig. \ref{fig:mhalo_mstar+gals} shows that galaxies with low values of \mstar \ display substantial scatter in their \mhalo \ values, ranging from \mhalo$\approx 10^9 \; M_\odot$ to \mhalo$\approx 10^{12} \; M_\odot$. The scatter in Fig. \ref{fig:mhalo_mstar+gals} cannot be easily attributed to any specific galactic property; this stands in contrast to the situation regarding the DC14 analysis in the \vrot-\vhalo \ plane (bottom panel of Fig. \ref{fig:vhalo_vrot+gals}), where the physical origin of the scatter is well understood in terms of the range in \Rout \ values. 
Regardless of the magnitude of the scatter however, Fig. \ref{fig:mhalo_mstar+gals} demonstrates once again that reionization has no impact on the assessment of the TBTF problem for field dwarfs; the interpretation of Fig. \ref{fig:mhalo_mstar+gals} does not depend on whether the \citet{Brook2014} or \citet{Sawala2015} AM relation is considered.

\begin{figure}
\centering
\includegraphics[width=\columnwidth]{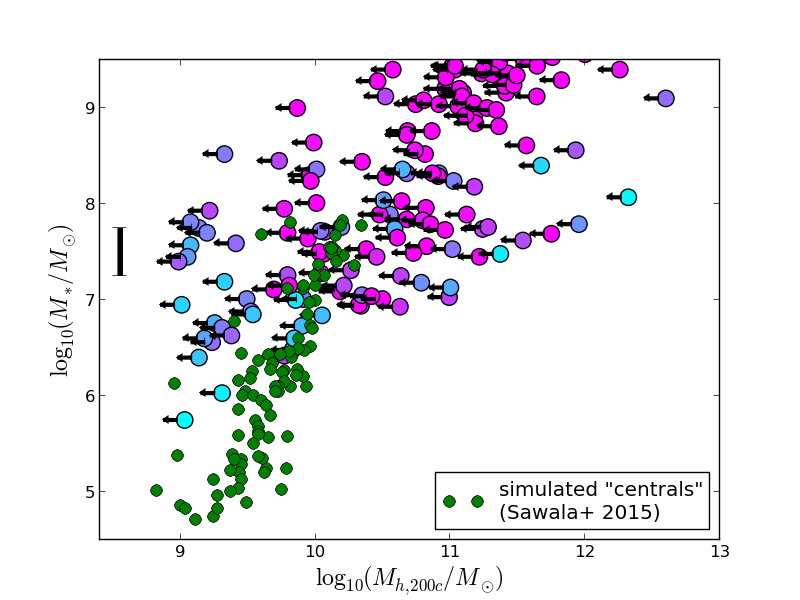}
\caption{Similar to Fig. \ref{fig:mhalo_mstar+gals}, but including only the low stellar mass portion of the plot. The green dots represent simulated ``central'' galaxies from the hydrodynamic simulations presented in \citet{Sawala2015}. Please refer to Appendix \ref{sec:appendix_a} for the scientific interpretation of this figure. }
\label{fig:gals_vs_centrals}
\end{figure}

The fact that the \mstar-\mhalo \ relation of dwarf galaxies displays large scatter has been already pointed out in the literature. For example, \citet{Garrison2014} find that there is no correlation between \mstar \ and \mhalo \ in low-mass isolated dwarfs in the Local Group (see their Fig. 12). \citet{Oman2016} also find large scatter in the \mstar-\mhalo \ relation of field dwarfs. On the other hand, this behavior is not present in hydrodynamic simulations. For example, Fig. \ref{fig:gals_vs_centrals} compares simulated ``centrals'' from \citet{Sawala2015} with our data, illustrating the much tighter \mstar-\mhalo \ relation obtained for simulated dwarfs (see also much more detailed discussion in \citealp{Oman2016}).     
In general, the issue of poor correlation between \mstar \ and \mhalo \ for dwarf galaxies may represent an obstacle for the assessment of the TBTF problem based on stellar mass abundance matching. This is because the AM method itself is based on the assumption of a monotonic and relatively tight relation between the two quantities of interest. 
It is therefore unclear how to interpret the comparison between the average relations derived by AM and the individual galactic datapoints. In that respect, the much better correlation seen in the corresponding NFW analysis in the \vrot-\vhalo \ plane (top panel of Fig. \ref{fig:vhalo_vrot+gals}), may offer an advantage for studies based on the galactic velocity function.

\section{Scatter in halo concentrations}
\label{sec:appendix_b}

In \S\ref{sec:kinematic_analysis}, the internal kinematics of galaxies are analysed based on mean concentration density profiles for the host halos. In reality, cosmological simulations predict that halo concentrations have a lognormal scatter about the mean mass-concentration relation of $\sigma_{\log_{10}c} \approx 0.1$ dex \citep[e.g.,][]{DuttonMaccio2014}. Concentration scatter is important, because the constraint that can be placed on \Vhalo \ based on one kinematic datapoint depends on the assumed concentration of the halo profile. In general, a density profile of lower than average concentration will lead to a less stringent constraint on \Vhalo, while a density profile of higher than average concentration will lead to a more stringent constraint.

\begin{figure*}
\centering
\includegraphics[width=\linewidth]{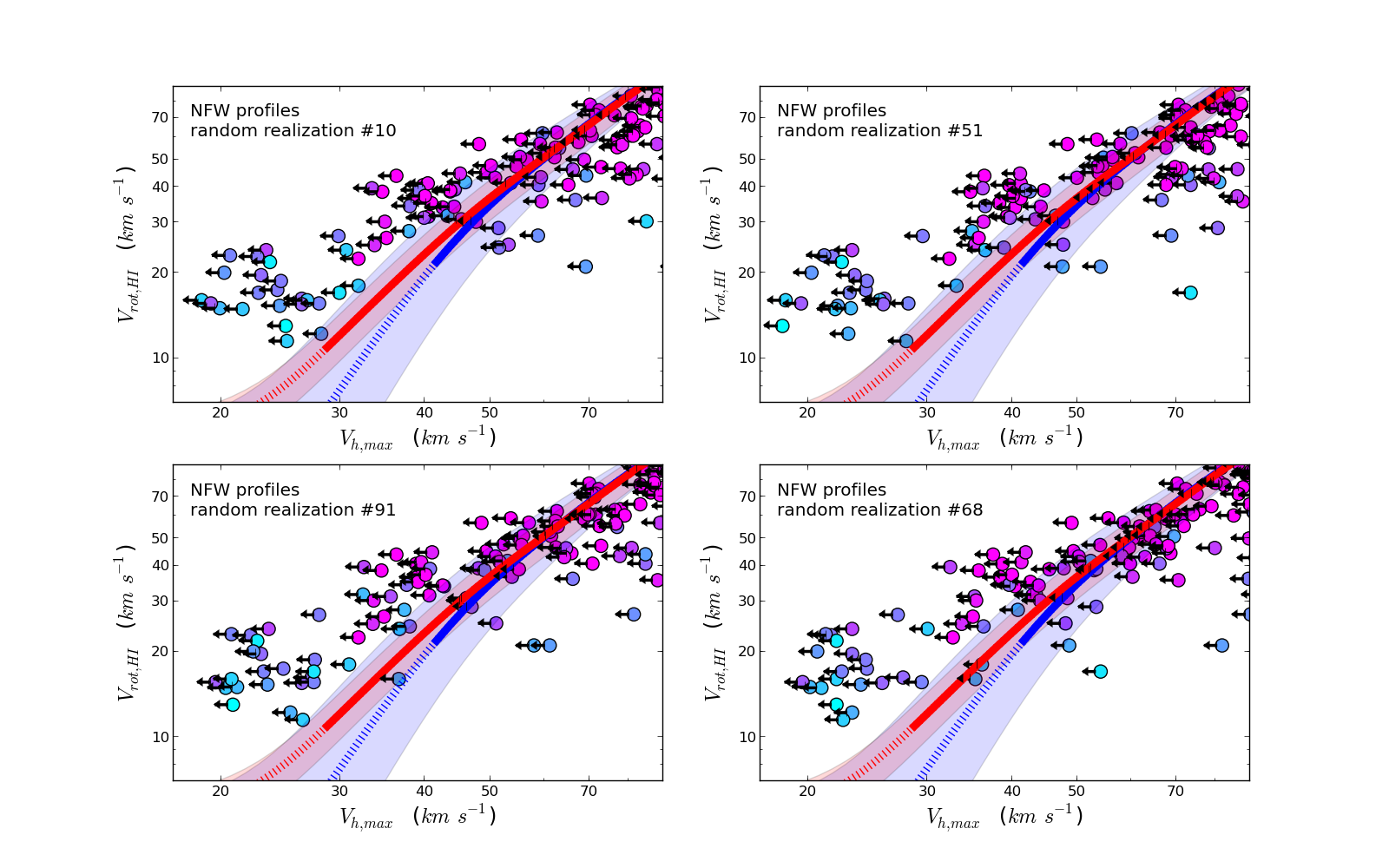}
\caption{Same as the top panel of Fig. \ref{fig:vhalo_vrot+gals}, but including scatter in the concentration of NFW halo profiles. Each panel corresponds to one out of a hundred random realizations, whereby the host halo of each galaxy is assigned a random value of concentration drawn from a lognormal distribution with $\sigma_{\log_{10}c} =0.1$ dex, centered on the mean mass-concentration relation in a \textit{Planck} cosmology \citep{DuttonMaccio2014}.}
\label{fig:concentration_scatter_cdm}
\end{figure*}

In order to check whether concentration scatter affects the results presented in Fig. \ref{fig:vhalo_vrot+gals}, we repeat here the kinematic analysis of \S\ref{sec:kinematic_analysis} but each time assigning to the host halos of our galaxies random values of concentration according to the lognormal scatter about the mean mass-concentration relation. The random concentration values are assumed to be uncorrelated among different objects, and independent of all properties of the galaxy other than the host halo mass. 

Figure \ref{fig:concentration_scatter_cdm} shows four realizations of the NFW kinematic analysis including concentration scatter, out of a total of one hundred performed. By comparing Fig. \ref{fig:concentration_scatter_cdm} with the top panel of Fig. \ref{fig:vhalo_vrot+gals} it is clear that the inclusion of concentration scatter in the NFW kinematic analysis does not change the results reached in Sec. \ref{sec:results}. We have visually inspected all one hundred random realizations, and find none where the TBTF problem has been resolved.

\begin{figure*}
\centering
\includegraphics[width=\linewidth]{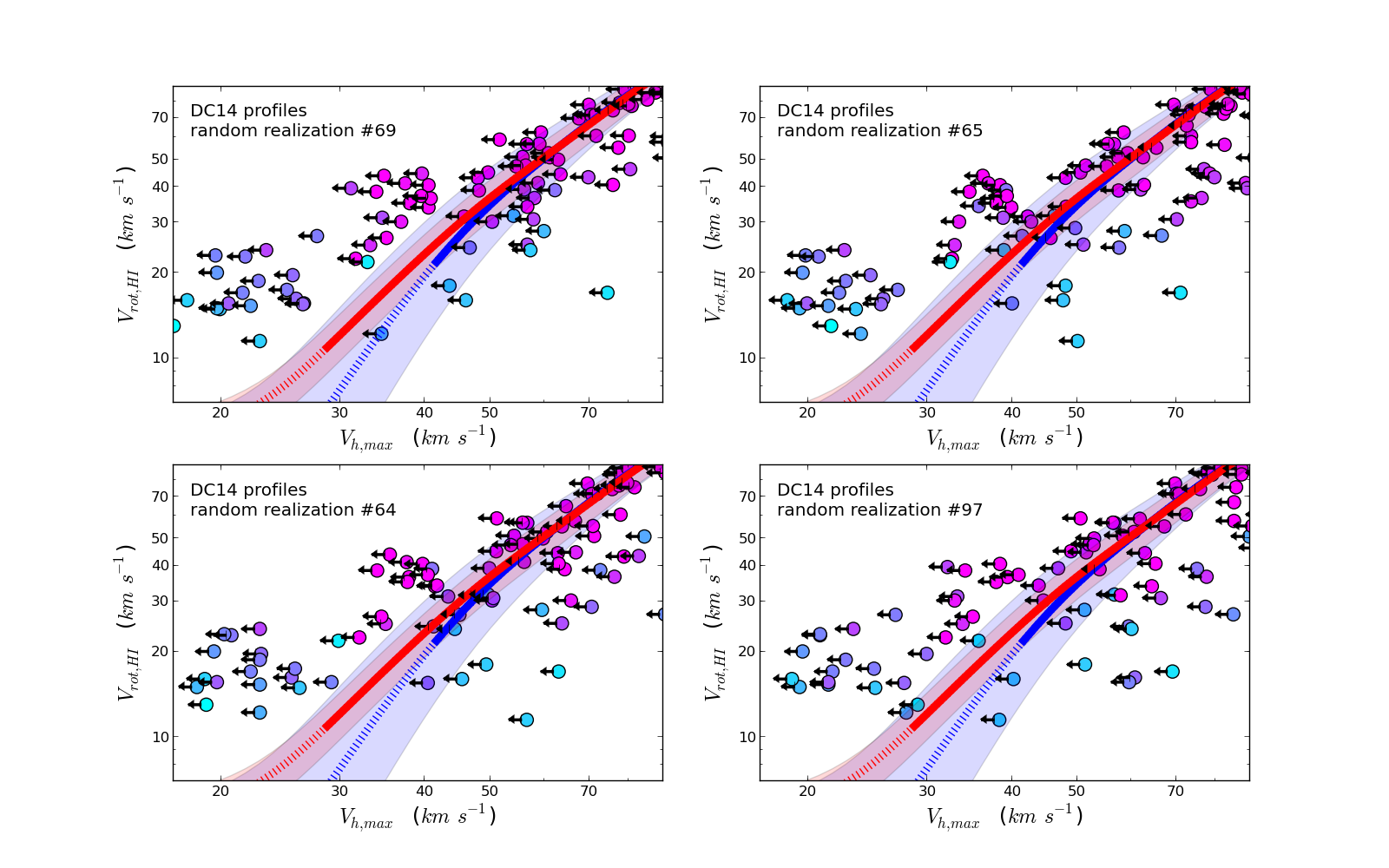}
\caption{Same as Fig. \ref{fig:concentration_scatter_cdm}, but concerning the DC14 kinematic analysis (in correspondence with the bottom panel of Fig. \ref{fig:vhalo_vrot+gals}).}
\label{fig:concentration_scatter_dc14}
\end{figure*}

Figure \ref{fig:concentration_scatter_dc14} illustrates the results of including concentration scatter in the DC14 kinematic analysis. Keep in mind that final concentration of a DC14 profile may include a baryonic correction, in the case of high stellar-to-halo mass ratio galaxies. In our analysis, the scatter is applied to the initial (i.e., DM-only) concentration parameter. Both in Fig. \ref{fig:concentration_scatter_dc14} and in the bottom panel of Fig. \ref{fig:vhalo_vrot+gals} we see that the galaxies which are compatible with the AM relations are most often the ones with low values of \Rout. As discussed in \S\ref{sec:kinematic_analysis} these galaxies have little constraining power on the \Vhalo \ of their host halo, when a cored profile is assumed. If we restrict instead ourselves to datapoints with \Rout$> 1.5$ kpc, then there are eight realizations out of the one hundred (8\% of cases) where the TBTF problem seems to have been resolved.

Overall, we conclude that the inclusion of scatter in halo concentrations does not significantly change the conclusions reached in Sec. \ref{sec:results} or the discussion in Sec. \ref{sec:discussion}. This holds certainly true for the NFW analysis and almost certainly true for the DC14 analysis. On the other hand, if the galaxies in our sample with \Vrot $\lesssim 25 - 30$ \kms \ were to be hosted by halos with systematically lower than average concentrations, then the TBTF problem could be alleviated or perhaps even resolved.

\end{document}